# Temperature effects in deep-water hydrate foam

Alexander V. Egorov[a,1], Robert I. Nigmatulin[a], Aleksey N. Rozhkov[a,b]

[a]*P.P. Shirshov Institute of Oceanology, Russian Academy of Sciences, 36 Nahimovski prospect, Moscow 117997, Russia*

[b]*A. Ishlinsky Institute for Problems in Mechanics, Russian Academy of Sciences, 101(k.1) Prospect Vernadskogo, Moscow 119526, Russia*

## ABSTRACT

This study focuses on heat and mass exchange processes in hydrate foam during its formation from methane bubbles in gas hydrate stability zone (GHSZ) of the Lake Baikal and following delivery of it in open container to the lake surface. The foam was formed as a result of methane bubble collection with a trap/container. The trap was inverted glass beaker of diameter of 70 mm and 360 mm long. Open bottom end of the beaker used as enter for bubbles ascended from the lakebed. At a depth of 1400 m all bubbles which fed to the trap were transformed here into solid hydrate foam. The sensitive thermometer was mounted in the middle of the trap and recorded the temperature inside trap. The fate of the bubbles in the trap was recorded by video-camera. Hydrate front, i.e. the border between forming hydrate foam and pure water, propagated slowly from the top to the bottom of the trap during bubbles collection. The temperature in the trap was slightly larger (by 0.05 $^{o}$C) than the temperature of ambient water. However as soon as the front crossed the location of temperature sensor, the temperature jump by ~+1.1 $^{o}$C was recorded. After that during 3 hours the temperature exponentially relaxes to an asymptotic magnitude which however exceeded the temperature of ambient water by 0.3 $^{o}$C. Thus hydrate foam formation occurs with heat dissipation with maximal intensity on hydrate front. The dissipation does not stop within hydrate volume at the constant depth of 1400 m during at least of 3 hours. After bubbles collection the trap with the foam was ascended from the lakebed to the surface, initially within GHSZ and then above GHSZ. During ascend within GHSZ with velocity about 0.375 m/s two remarkable features were observed. First one is displacement of the water from the trap by gas which was released from the foam sample under the decrease of the hydrostatic pressure. It was found that gas expansion follows to Boyle-Mariotte law in which initial volume of gas is equal approximately to volume of the foam sample. The second feature is continuous decrease of the temperature in the foam up to a level of negative magnitude in a depth interval of 1400 - 750 meters. Above 750 m temperature decrease was changed by small growth. However once the trap ascended above top boundary of GHSZ at a depth of 380 m, the temperature fell sharply. Falling to a negative values -0.25 $^{o}$C, the temperature sharply stabilized and did not changed further until the trap reached the surface. The decreasing of the temperature during the ascent is due to the cooling of gas as a result of the performing of the work against the forces of hydrostatic pressure. The temperature drop at the boundary of the GHSZ is due to the absorption of heat by the decomposition of hydrate. The keeping of the foam temperature at the constant level -0.25 $^{o}$C above GHSZ reduces hydrate decomposition and is exhibition of effect of self-conservation of hydrate.

[1] Corresponding author. P.P. Shirshov Institute of Oceanology, Russian Academy of Sciences, 36 Nahimovski prospect, Moscow 117997, Russia

*E-mail address*: avegorov@ocean.ru (A.V. Egorov)

# 1. Introduction

The formation of hydrate matter from methane within gas hydrate stability zone (GHSZ) and decomposition of this matter above GHSZ are components of global control of the movement of methane worldwide. Transfer of methane from the lakebed to the hydrosphere and from the hydrosphere to the atmosphere can influence climate change due to the huge flux of methane into the atmosphere from unstable methane hydrate deposits in oceanic shelf in case of a global temperature elevation (Judd et al., 2002; MacDonald et al., 2002). One possible mechanism of methane transfer is ascent of methane bubbles from the deep-water methane deposit. Within GHSZ methane in the bubbles can be transformed into hydrate matter such as single bubbles with hydrate envelope (MacDonald et al., 2002; Ribeiro and Lage, 2008; Leifer and Culling, 2010; Li and Huang, 2016), hydrate foam (Luo et al., 2007; Egorov et al., 2014) or hydrate granular matter (Egorov et al., 2010). Usually the phase transformations are accompanied by the effects of heat consumption or dissipation. Depending on external conditions heat effects can influence on the processes of the phase transformations.

The mechanics of the phase transformations of deep-water methane is crucial for the implementing the optimum and safest technology during underwater work on the bottom areas with methane hydrate deposits (Ribeiro and Lage, 2008; Hagerty and Ramseur, 2010). Heat phenomena are also important for the development of the technology of gas recovery from the deep-water hydrate matter - bubbles, granular matter or solid foam.

The basic experiments with deep-water gas hydrates were conducted in Lake Baikal (Sagalevich and Rimskiy-Korsakov, 2009; Sagalevich, 2011). Large deposit of methane hydrate was discovered in deep mud volcano ‘St. Petersburg’ (Fig. 1) at a depth of 1400 m within GHSZ (Fig. 2). The hills of monolithic hydrate of 1 - 3 m height are located in the area of few hectares (Fig. 3). There are also seeps of methane in the form of bubbles that release from the sediment into the hydrosphere in certain places of deposit area (Fig. 4).

The samples (fragments) of hydrate were excavated from monolithic hydrate deposit by means of mechanical arm of manned submersible (MS) 'Mir'. The samples were placed in the non-sealed container and delivered to the surface (Egorov et al., 2011). The container had grid-like bottom, so the mass exchange between the container content and the ambient media could be possible only through grid-like bottom. Above GHSZ hydrate decomposition provided filling of the container by gas that promoted to hydrate conservation in the course of hydrate transport to the surface.

To study the details of heat effects, the container was equipped by two thermometers, one of which was mounted under the top cover of the container and the second was mounted on the container bottom (Egorov et al., 2016). No changes were observed in hydrate samples during the ascent within GHSZ. The water temperature in the container remained the same as one of the ambient water ~+3.5°C. However as soon as the container crossed the upper bounder of GHSZ, the first signs of decomposition of hydrate and its transformation into the free methane gas occurred. Gas filled the container and displaced water from it. At a depth of 300 m the upper and lower thermometers in the container recorded simultaneously noticeable decrease of temperature. The temperature in the upper part of container decreased to -0.25°C at a depth of about 200 m. After that the temperature remained constant until the lake surface was reached. The temperature at the bottom of container reached -0.25°C at a depth of about 100 m and after that it did not vary during further ascent. The observed effects are explained by the formation of gas phase in the container and an ice layer on hydrate surface caused by heat consumption during hydrate decomposition (self-conservation effect).

However fragments of monolithic hydrate are not sole hydrate matter, which was found in a deep mud volcano ‘St. Petersburg’. Another hydrate matter is solid hydrate foam which was formed in the container as a result of the collection of methane bubbles (Egorov et al., 2014). Similar to the container for hydrate fragments, the container for hydrate foam allowed mass exchange only through the bottom. The containers with hydrate fragment and hydrate foam were subjected to ascent with the same rate. While hydrate fragment was stable during the ascent up to the top boundary of GHSZ, gas from hydrate foam began to escape just after the beginning of the ascent. Gas displaced the water from the container with the ascent of the container. The record of the trajectory of gas-water meniscus in the container displayed that gas expands in the container with the decrease of ambient hydrostatic pressure almost according to the Boyle–Mariotte law in which initial gas volume is equal to the total volume of the foam formed at the initial depth. Numerical analysis of the meniscus trajectory shows very high foam porosity and the conservation of the mass of gas in the trap during its ascent.

It should be noted that in another thermobaric conditions, namely in the area 'Gorevoy Utes' at a depth of 860 m we observed methane bubbles transformation into hydrate granular matter, but not into the consolidated foam (Egorov et al., 2010, 2014).

This study was undertaken to assess heat and mass transfer in the processes of the transformation of deep-water methane bubbles into solid hydrate foam and vice versa. The main purpose was to answer the question regarding what are the features of heat effects in high porous hydrate foam during it ascent from the lakebed to the basin surface and what is difference from heat effects in non-sealed containers with monolithic hydrate fragments.. Another point of the research is heat effects in hydrate foam during its formation from deep-water methane bubbles.

# 2. Regional setting

Deep-water experiments with methane bubbles and their transformation in hydrate foam were conducted in the vicinity of the shallow deposit of methane hydrate in the area 'St. Petersburg' (Figs. 1-4). during the expedition of the Russian Academy of Sciences 'MIRI NA BAIKALE 2008-2010' (Sagalevich and Rimskiy-Korsakov, 2009; Sagalevich, 2011; Egorov et al., 2014). The 'St. Petersburg' area is a deep mud volcano named 'St. Petersburg'. A deep mud volcano is preferable locality for hydrate deposit formation (Egorov and Rozhkov, 2010). The coordinates of the volcano 'St. Petersburg' are 53°52.97 N, 107°09.99 E. A depth here is 1400±10 m. The main part of the water in Lake Baikal including 'St. Petersburg' area maintains a low temperature (+3.2 °C to +3.5 °C) throughout all of the seasons. The water in the lake is subjected to high hydrostatic pressure due to the large depth. Thus GHSZ occupies the main part of the water in the 'St. Petersburg' area. It was found that the upper boundary of the GHSZ in Lake Baikal is located at a depth of $z_S$~380 m (Egorov et al., 2010), that compares well with the estimate of Katz, 1959 shown in Fig. 2. Another remarkable feature of the 'St. Petersburg' area consists in the presence of many seeps of methane in the form of the bubbles that escape from the sediment to the hydrosphere. Taking into account all these peculiarities as well as the convenient geographic location of the 'St. Petersburg' area, one can conclude that this area in Lake Baikal can be considered as a superior natural laboratory for study of deep-water methane hydrate.

The dive for study heat effects in the lifting methane bubbles was conducted at August 9, 2010. The similar heat study with the fragments of the monolithic hydrate (Egorov et al., 2016) was conducted here two days later at 11 August, 2010.

# 3. Materials and methods

The experiments were conducted with the Mir deep-water manned submersibles (MS). To provide a collection of the deep-water methane bubbles near the lakebed, the observation of their transformation into hydrate matter, the ascent of the forming hydrate matter from the lakebed to the surface, all with simultaneous measurement of the temperature inside the trap, a special trap/container 'Thermo' was created. The basic element of the trap 'Thermo' looks as inverted glass beaker: the cylindrical Plexiglas tube of size $\phi 70 \times 360$ mm with the top closed by the Plexiglas cover (Fig. 5). To speed up the collection of the bubbles a transparent plastic funnel was attached to the lower open end of the tube.

To measure temperature history in the trap a sensitive thermometer was mounted inside the trap. The thermometer was manufactured by THK nke INSTRUMENTATION (France). The thermometer is designed for the deep-water measuring of the temperature in the hydrosphere or sediment. The sensitivity of the temperature sensor is in the temperature range $0 \div 25°$ C is 0.003 °C. The measuring element (sensor) is located in the middle of the thin pin (Fig. 5). The pin length is 126 mm. Temperature measurements were carried out during the dive at intervals of 1 minute. Data are recorded in the thermometer memory. After dive the data are downloaded to the computer using the remote radio reader. The same thermometer was used in similar experiments with the fragments of the monolithic hydrate (Egorov et al., 2016).

Before the dive the trap was placed in the bunker of MS 'Mir' as it is shown in Fig. 6. The mechanical arm of submersible 'Mir' took the trap from the bunker when the deep-water bubble collection began.

To collect the methane bubbles the mechanical arm of the submersible 'Mir' arranged the trap 'Thermo' near the lakebed in the places of intensive methane bubbles release. The bubbles fed to the trap through open lower end of the trap (Fig. 7).

After end of the bubble collection the trap with forming inside hydrate matter was delivered to the lake surface by submersible 'Mir'. During the ascent the trap remained in the mechanical arm without change of the orientation. The ascent was carried out with approximately constant rate $u_0$=0.375 m/s.

At all stages of the dive the trap content was monitored with the video-camera.

# 4. Results and discussion

*4.1. Bubbles collection near the lakebed*

When the methane bubbles entered into the trap they transformed into hydrate foam (Fig. 7). The hydrate envelope prevented the bubble coalescence and the column of the bubbles filled the trap forming hydrate foam. However just after the delay of the bubbles by the trap the bubble's envelope remains thin and flexible. The bubbles at bottom part of the column can be easily deformed or shifted by trap shaking. Such foam behavior reminds the behavior of the conventional foam during its motion in the channel (Bazilevsky and Rozhkov, 2012). In this stage the bubbles are subjected to the packing in the trap due to the competition of the bubble buoyancy and the surface tension. Meanwhile the hydrating of the bubbles causes the growth of hydrate envelope thickness and correspondent hardening of the foam with time. The observations show that few tens of minutes after the formation the foam becomes solid. On the qualitative level we suppose the structure of the foam in the process of its formation with temporal 'islands' of water between the bubbles as shown in Fig. 8. Obviously, the real structure depends on the rate of the bubbles flux to the trap because it defines the competition of the water drainage from the foam and the bubble envelope hydrating.

Figure 9 presents the result of the temperature measurement of the trap content in the process of the bubble collection $T(t)$ and the trajectory of the propagation of the front of the foam formation $x(t)$. The front coordinate $x$ is distance between the top of the pin and the front position (Fig. 5). The trajectory was obtained by means of the video-data processing.

The hydrate formation accompanies by heat dissipation. As soon as the first bubbles are entered to the trap, their hydrating caused slight growth of the temperature in the trap approximately from 3.142 $^{o}$C up to 3.223 $^{o}$C. However a significant temperature jump was recorded only in the moment $t$=17:34:26 when the foam front reached the position of the temperature sensor. The temperature was elevated from 3.223 $^{o}$C up to 4.295 $^{o}$C during 18 minutes. Note that this time compares well with the response time of thermometer ~15 min. After reaching the maximum the temperature relaxes very slowly during 3 hours in spite of the stop of the bubbles collection just 25 minutes after the beginning of the temperature jump. The temperature decrease can be approximated by the exponential function $T=(T_m-T_i)\exp(-(t-t_m)/\theta)+T_i$, where $T_m$=4.3305 $^{o}$C, $T_i$=3.45937 $^{o}$C, $t-t_m$ is time [seconds] passed after reaching the maximal temperature, $\theta$=1953.4666 s (Fig. 10). Asymptotic temperature $T_1$=3.45937 $^{o}$C remains noticeably larger than the temperature of the ambient water 3.142 $^{o}$C.

Elevated temperature in the trap during 3 hours can be explained by the low heat exchange with the ambient media. Indeed, as it follows from heat equation the time interval of order of $\Delta t\sim(\Delta x)^2/\alpha$ is required to conduct the temperature pulse through the distance $\Delta x$. The constitutive parameter $\alpha$ is thermal diffusivity (m$^2$/s): $\alpha=k/\rho c_p$, where $k$ is thermal conductivity (W/(m·K)), $\rho$ is material density (kg/m$^3$), $c_p$ is specific heat capacity (J/(kg·K)). As it was found by Egorov et al. (2014) hydrate foam in the trap is characterized by very high magnitude of the porosity. In this case the main conductor of the temperature is methane gas. The thermal conductivity and the specific heat capacity slightly depend on the pressure and are equal to $k$=0.0543 W/(m·K) (Thermalinfo.ru, 2016a ) and $c_p$=3047 J/(kg·K) (Thermalinfo.ru, 2016b). The density of a methane at a depth of 1400 m is equal to $\rho$=98.73 kg/m$^3$ (Egorov et al., 2010). Thus the magnitude of the thermal diffusivity of the methane at a depth of 1400 m can be evaluated as $\alpha=1.8050\cdot10^{-7}$ m$^2$/s. Taking the characteristic length of heat exchange process the radius of the glass beaker ($\Delta x$)=0.035 m, we obtain the characteristic time of the temperature exchange process in the trap $\Delta t\sim(\Delta x)^2/\alpha$=6786.7 s. The order of this value compares satisfactory with the order of the temperature relaxation time $\theta$=1953.467 s. in the plot of Fig. 9.

The observed features show that heat dissipation due to the hydrating process is maximal at the front of the foam formation. However it is possible to suggest that the hydrating does not stop inside the foam after the propagation of the foam formation front because the asymptotic temperature exceeds the temperature of the ambient water. The continuation of the hydrating can be explained by the participation in the hydrating process of the water which was stored in the 'water island' inside the foam (Fig. 8).

Note that at the level of the lake surface mentioned above thermal properties are: $k$=0.0304 W/(m·K), $c_p$=2290 J/(kg·K), $\rho$=0.7168 kg/m$^3$, $\alpha=1.852\cdot10^{-5}$ m$^2$/s, $\Delta t\sim(\Delta x)^2/\alpha$=66.1447 s. Thus the rate of the temperature exchange in the trap increases with the decrease of the depth.

The temperature in the trap 'Thermo' fell sharply when MS 'Mir' began the ascent.

### *4.2. Bubbles transport from the lakebed to the top bounder of GHSZ*

At 20:37:00 (2:24:00 after of the end of bubble collection), the MS 'Mir' began the ascent of the trap 'Thermo' with hydrate foam. During the ascent from the lakebed to the surface, the current depth, the temperature in the trap, the temperature of the ambient water and the video-image of the trap content were recorded.

The variation of the temperature in the trap during the ascent and the trajectory of the ascent are shown in Fig. 10. In the plots all data for hydrate foam are compared with the similar data for the monolithic hydrate, obtained in the dive two days later (Egorov et al., 2016). In the latter case the fragments of the monolithic hydrate were placed in the container 'TV-box'. The container 'TV-box' was a transparent parallelepiped of the size of 297×210×210 mm. The bottom wall was absent and the monolithic hydrates were stuffed into the container through the bottom due to hydrate positive buoyancy. The temperature in the container was recorded by two thermometers which were arranged in the top and the bottom parts of the container.

The plots show a great difference between the temperature variations in the trap 'Thermo' with hydrate foam and in the container 'TV-box' with the monolithic hydrate during the ascent within GHSZ.

During the ascent the temperature in the trap 'Thermo' fell significantly and reached the minimal level of -0.41691 $^{o}$C at a depth of 750 m (21:06:26). During the ascent above this depth the temperature elevated. The temperature growth up to +1.0448 $^{o}$C was recorded at a depth of 292 m (21:26:26). This depth 292 m coincides well with the top boundary of GHSZ at this temperature (Fig. 10).

In spite of a very similar trajectories of the ascent of the trap 'Thermo' and the container 'TV-box', the top and the bottom thermometers in the container 'TV-box' showed very low temperature variation within GHSZ. The temperature in the container was of order of the temperature of the ambient water, i.e. about +3.5 $^{o}$C. Approximately the same temperature was recorded by external thermometer of MS 'Mir'.

The high variation of the temperature in the trap 'Thermo' becomes clear from the observation of the video-record of the trap content during the ascent. The video-images demonstrate the displacement of the water from the trap by means of the methane gas, which released from hydrate foam during it ascent (Fig. 11). As soon as the trap began to ascend the ambient hydrostatic pressure decreased. The pressure fall caused gas expansion in the foam. As it was found in the numerous experiments with the foam samples in different traps (Egorov et al., 2014) gas released from the foam without visible deformation of foam structure. The video-records of gas-water meniscus trajectories validated certainly that gas expansion in the traps follows to Boyle-Mariotte law in which initial volume of gas is equal approximately to the volume of the foam sample. Analysis shows that Boyle-Mariotte law is obeyed if solid part of the foam is formed by very thin hydrate sheets. Probably sheets cannot keep pressure drop between gas inside foam and ambient hydrostatic pressure. It is not clear so far what kind of foam distortion happens, but gas free motion through the foam certainly occurs.

Gas expansion performs the mechanical work against the hydrostatic pressure. The performance of the work causes gas cooling. However, the process is far from being adiabatic. Indeed, the dash curve '*A*' in Fig. 10 shows the variation of the temperature with a depth for the adiabatic process, which is described by Poisson equation

$$T=T_0(z/z_0)^{1/\kappa-1}, \qquad (1)$$

where $T$ is current absolute gas temperature, $T_0$=276 $^{o}$K is absolute temperature of gas at the moment of the start of MS ascent from the bottom, $z$ is current depth, $z_0$=1400 m is depth of the start of MS ascent, κ=1.31 is adiabatic index for methane. Obviously, the adiabatic description in Fig. 10 is not consistent with the observations. The internal energy of gas is consumed by the performance of the work, but heat loss is immediately compensated by heat flux from the environment through the trap wall and gas-water meniscus. The decrease of the temperature in the trap is small and it is controlled by an extremely small difference between heat loss due to the performance of work and heat coming from the environment.

Obviously, the certain differences between the data of thermometers T1, T2 and *Mir* in Fig. 10 are caused also by the work performance by a small gas volume which was entered into the container 'TV-box' during the collection of the fragments of the monolithic hydrates (Egorov et al., 2016).

An unexpected result is appearance of a minimum at 750 m in the temperature dependence versus the depth. Two reasons can provide the temperature minimum. The first one is growth of the thermal diffusivity with the decreasing of a depth as it is described in section 4.1. The growth of the thermal diffusivity causes the intensification of heat exchange between the trap content and environment that leads to equalization of temperatures inside and outside the trap.

The second reason of the temperature minimum is progressive increase (faster than linear) of the surface of gas bubble in the trap. The ascent with constant velocity is assumed. This factor is considered in Appendix (see also Egorov et al., 2012) within the framework of heat balance equation, which accounts the competition of heat loss through the surface of gas bubble in the trap and heat removal by work performance. It was demonstrated that minimum on the curve $T=T(Z)$, $Z=z/z_0$ appears if bubble surface $S=s/s_0$ (ratio of current and initial surfaces of gas bubble) varies with a depth $Z$ faster than $Z^{1}$.

By processing of the video-record of the ascent of the trap 'Thermo' the trajectory of the meniscus in the trap was calculated. Based on these data, the dependence of gas volume versus depth was found. It is presented in Fig. 15 of Ref. (Egorov et al., 2014). The data for the trajectory of the meniscus allowed also to recover similar relationship for gas surface in the trap $S(Z)$. Figure 12 shows the desired dependence by means of diamonds. Points were calculated with account the shape of the trap (cylindrical part + funnel) and the presence of the thermometer in the trap.

Until the meniscus is in a cylindrical part of the trap, the surface varies approximately inversely with the depth: $S=Z^{1}$ as it is necessary for monotonic decrease of the temperature. As soon as the meniscus comes to expanding part of the trap (the funnel), the surface of gas bubble begins to grow much faster - Figs. 11, 12. Video frames show that the transition of the meniscus in the funnel occurred at a depth of about $z_1$=800 m ($Z_1$=0.5714). Around this depth the deviation from the dependency $S=Z^{1}$ is noted. The approximation $S=Z^{1}+109.4(1-Z)^{9.6088}$ describes experimental points (Fig. 12). Using of this relation in numerical calculations leads to a minimum on the temperature curve (Appendix, Egorov et al., 2012). Indeed, the exit of the meniscus into the funnel is equivalent to an appearance of an additional radiator, which intensifies heat exchange and equalizes the temperature inside and outside the trap.

Numerical calculations of Ref. (Egorov et al., 2012) also show that the magnitude of temperature fall in the trap increases with the increase of the ascent velocity. It means that by increasing of ascent velocity it is possible to enter in GHSZ at a depth shallower than top bounder of GHSZ in Baikal 380 m (Fig. 10 A). Therefore the transport of hydrate foam to the surface with more high velocity is method to decrease hydrate decomposition during its way to the surface due to more late enter into GHSZ. Obviously this is one way to improve the efficiency of gas hydrate recovery technology.

*4.3. Bubbles transport from the top bounder of GHSZ to the surface*

As soon as the trap with hydrate foam crossed the top bounder of GHSZ, a sharp jump of the temperature from +1.0448 °C at a depth of 292 m up to -0.4997 °C at a depth of 241.28 m was observed (Figs. 10 C). The average value of the temperature in a depth range 241.28 ÷ 0 m was -0.2567±0.0345 $^0$C with the standard deviation 0.11962 $^0$C.

These data compare well with the data of temperature measurements in the container 'TV-box'. Both thermometers in the container 'TV-box' also recorded the decrease of the temperature after depth of 300 m up to the level -0.25 $^0$C, but it was not as sharp as in the case of thermometer 'Thermo' (Fig. 10 C). The top thermometer T1 displayed the average temperature -0.2214±0.0060 $^0$C, with the standard deviation of 0.0189 $^0$C in a depth range 144 ÷ 0 m. The bottom thermometer T2 displayed the average value of the temperature -0.2599± 0.0034 $^0$C, with the standard deviation of 0.0215 $^0$C in a depth range 49 ÷ 0 m. As it was mentioned in (Egorov et al., 2016) the time delay in data of bottom thermometer T2 is related with finite time during which gas phase reached thermometer T2 in the container 'TV-box'.

Once MS 'Mir' reached the lake surface, the temperatures in the trap 'Thermo' and in the container 'TV-box' began to grow rapidly due to relatively high temperature of the surrounding water (~10 °C) and the stop of heat consumption due to gas expansion.

It is assumed that at a depth less than top boundary of GHSZ methane hydrate decomposition begins. Our previous measurements determined the top boundary of GHSZ in Lake Baikal at a depth of about 380 m (Egorov et al., 2014). The boundary can be shallower if the temperature around hydrate is lesser than +3.5 $^{o}$C (curve 'GHSZ' in the plots of Fig. 10). Therefore one can suggest that the temperature changes at a depth less than 300 m are caused by hydrate decomposition in the thermobaric conditions which cannot provide hydrate stability. Hydrate decomposition occurs with the absorption of heat, which decreases the temperature of trap or container content. On the other hand the water releases due to hydrate decomposition and it covers hydrate surface. The cooling to a negative Celsius temperature causes the freezing of the water and hydrate surface becomes covered by ice envelope, which probably prevents hydrate decomposition. This is well-known effect of self-preservation of hydrate (Davidson et al., 1986; Istomin and Yakushev, 1992; Uchida et al., 2011). Apparently, then the system adjusts itself and maintains the constant temperature at the level of about -0.25 $^{o}$C. Once formed slight thawing accelerates the decomposition and consequent cooling returns the system to the equilibrium temperature level. On the other hand too low temperature forms a lot of ice which retards the decomposition and also returns the system to the equilibrium temperature level.

However our analysis of the strength state of the ice envelope around hydrate shows that in our experiments with the ascent of monolithic hydrates the ice envelope was too thin to keep inside the pressure which able to prevent hydrate decomposition (Egorov et al., 2016). Below we present the modification of model of self-preservation of hydrate which is based on the hypothesis of local blockade of methane microbubble growth in hydrate microfracture by means of ice microjam.

Concluding the section, we note that high water pressure lowers water freezing point (Wikipedia.org, 2016). However this effect is too small to reach the level of the temperature of -0.25 $^{o}$C as it is demonstrated in Fig. 10 C.

*4.4. Mechanical model of self-preservation of hydrate*

The approach is similar to one, which is used in the modelling of the processes of boiling, cavitation, rupture of liquid, etc. The basic assumption consists in the statement that the liquid volume losses its continuity due to the catastrophic growth of micro-bubbles when the pressure inside the micro-bubble exceeds the sum of Laplace pressure and the pressure of ambient media (Kornfeld, 1952; Knapp et al., 1970). Thus, three factors define the loss of continuity. The first factor, pressure inside the bubble is formed by the processes of gas release from the contact surface of bubble gas with another phase (solid or liquid). Gas release is controlled by liquid/solid nature and thermodynamics circumstances. The second factor, the micro-bubbles play the role of material defects in solid, which define the strength of solid materials. Micro-bubbles can remain in the material from all possible previous events or can be formed in the liquid as a result of the fluctuations (Kornfeld, 1952). And finally the third factor, pressure of ambient media can be controlled by different manner, including artificial methods.

During the expedition ‘MIRI NA BAIKALE 2008-2010’ we attempted to deliver hydrate fragment to the surface keeping it in mechanical arm of MS 'Mir' as shown in Fig. 5 of Ref. (Egorov et al., 2016). All attempts were failed because above GHSZ fragment destructed. However we could observe the processes preceding the destruction. Initially micro-bubbles nucleated in hydrate sample as demonstrated in Fig. 13. Bubbles formed in a lot of fractures in hydrate sample. Video-record shown that each fracture ejected the long sequence of tiny bubbles, which were captured by advancing water stream (Fig. 7 of Ref. (Egorov et al., 2016)). Thus, hydrate decomposition occurs through fractures where methane bubbles are nucleated.

Note that any element of flat surface of hydrate cannot be decomposed directly into methane and water even in the case of thermobaric instability. Indeed, let us suppose that decomposition occurs in water media on hydrate flat surface and bubble appears. The bubble will grows if Laplace pressure is less than pressure in the bubble. The pressure in bubble must be less than the critical pressure of hydrate stability, say $p_{380}$=38 MPa at the temperature 3 °C, to provide the decomposition. Therefore, the decomposition is possible if

$$2\gamma/r<p_{380}$$

where $\gamma$ is surface tension of the water, $r$ is bubble radius. Taking $\gamma$=0.0726 N/m, $p_{380}$=38 MPa, we obtain the critical radius of the bubble $r_c=2\gamma/p_{380}$=3.82 nm If there are no bubbles with such radius (or larger) on hydrate surface, then the decomposition is impossible here. The account of the pressure of an ambient media only increases the magnitude of critical radius $r_c$. The change of the water environment to gaseous atmosphere can assist to the decomposition if only a hydrate surface is dry. However it is unlikely due to high gas-hydrate wettability and presence of residual water and the water released from the partial hydrate decomposition.

To model self-conservation in gaseous atmosphere, we assume that initially micro-bubbles are available; but all of them are located in the fractures in hydrate (Fig. 14). As soon as the thermobaric conditions provide hydrate instability, the micro-bubble grows due to methane supply by hydrate decomposition process. In the water media the growth of the micro-bubble ejects gas from the fracture into the ambient water as it was observed in the experiments. The rest of gas in the fracture is used for nucleation of the next bubble and continuation of the bubbles ejection. The new water which was formed by hydrate decomposition mixes with ambient water and does not participate in hydrate decomposition more.

In gaseous atmosphere the new water remains in the fracture. The low heat exchange allows the water cooling up to freezing point due to heat dissipation in hydrate decomposition. Forming ice can serve as the jam which blockades gas exit from the fracture. Continuation of hydrate decomposition causes the pressure increase in the blockaded volume up to the critical level $p_{380}$ at which the decomposition ceases. Thus, given fracture with its content can be conserved.

The ice formation causes the decrease of the temperature in the trap (or container) up to the level, which is a little less than the water freeze temperature ~0 °C. The further decrease of the temperature is retarded by the cease of hydrate decomposition due to the more and more ice conservation of the fractures. Thus, self-adjustable system of hydrate and ambient gas media maintains the temperature in the trap (or container) on the constant level (-0.25 °C according to experiment) as a result of heat balance of heat flux from ambient media and heat consumption by hydrate decomposition.

Figure 14 shows the simplest scenario of the formation of the ice jam in the form of ice ball, which keeps the pressure in the bubble at the level of $p_{380}$. Obviously, other forms of ice jam are possible. Let us assume for simplicity that semi-spherical (~radius $r_0$) volume of hydrate was transformed into ice ball (spherical ice layer) of external radius $r_0$ and spherical bubble of methane of radius $r_1$. Hydrate mass is equal to $m_{gh}={}^1/_2\rho_{gh}{}^4/_3\pi r_0^3$, the ice mass is equal to $m_i=(1-\kappa)m_{gh}$, the ice volume is equal to $V_i=m_i/\rho_i$, where $\rho_{gh}$=900 kg/m$^3$ is hydrate density, $\kappa$=0.129 is the mass fraction of methane in hydrate, $\rho_i$=916 kg/m$^3$ is the ice density. Relation between radii $r_1$ and $r_0$ follows from the geometrical equation ${}^4/_3\pi(r_0^3-r_1^3)=V_i=(1-\kappa)\times{}^1/_2\,\rho_{gh}\,{}^4/_3\,\pi r_0^3/\rho_i$ as $r_1/r_0=(1-{}^1/_2\,(1-\kappa)\rho_{gh}/\rho_i)^{1/3}$=0.8302.

Correspondently the ice layer thickness is equal to $h$=0.1698$r_0$. It is remarkable that characteristics of the ice jam $r_1/r_0$ and $h/r_0$ have the magnitudes which do not depend on dimensions of the fracture and other characteristics of the hydrate breakup!

To break or not to break - that is the question, which arises when the ice ball is formed.

The inner stresses in the spherical ice layer $r_0$-$r_1$=$h$ are described by the exact solution of Lamé (Timoshenko and Goodier 1951, p. 359):

$$\sigma_t-\sigma_r=(3p_0(r_1/r_0)^3-p_{380}(r_1/r_0)^3(2(r_1/r_0)^3+1))/(2(r_1/r_0)^3((r_1/r_0)^3-1))+p_{380}$$

at the internal surface of the spherical layer, and

$$\sigma_t-\sigma_r=(p_0(2+(r_1/r_0)^3)-3p_{380}(r_1/r_0)^3)/(2((r_1/r_0)^3-1))+p_0$$

at the external surface of the layer. In these formulas, $\sigma_t$ is normal stress in a tangential direction, $\sigma_r$ is normal stress in a radial direction, $p_{380}$ is the pressure on the internal surface of the ice layer, $p_0$ is the pressure on the external surface of the ice layer (Fig. 14).

Taking $p_{380}$=3.8 MPa, we obtain:

at a depth of 300 m ($p_0$=3 MPa) $\sigma_t$-$\sigma_r$=2.8051 MPa (internal surface) and $\sigma_t$-$\sigma_r$=1.6051 MPa (external surface);

at a depth of 200 m ($p_0$=2 MPa) $\sigma_t$-$\sigma_r$=6.3114 MPa (internal surface) and $\sigma_t$-$\sigma_r$=3.6114 MPa (external surface);

at a depth of 100 m ($p_0$=1 MPa) $\sigma_t$-$\sigma_r$=9.8177 MPa (internal surface) and $\sigma_t$-$\sigma_r$=5.6177 MPa (external surface);

at a depth of 0 m ($p_0$=0.1 MPa) $\sigma_t$-$\sigma_r$=12.9734 MPa (internal surface) and $\sigma_t$-$\sigma_r$=7.4234 MPa (external surface).

According to the theory of the maximal shear stress of Tresca (Sedov, 1997) the difference between the normal stresses defines the destruction threshold in solid material. Thus to sustain the pressure inside the ice layer on the level of 3.8 MPa the strength of the ice must of the order of 1÷10 MPa.

On the other hand according to available experimental data (Petrovic, 2003), the strength of ice is of order of 1÷10 MPa. Therefore our estimates show that the order of the ice strength can occur sufficient to keep the pressure which provides gas hydrate stability in gaseous atmosphere. Thus it is not necessary to cover by the ice whole hydrate surface to suppress hydrate decomposition as it was discussed in (Egorov et al., 2016). It is sufficient to plug by the ice jams only most weak places of hydrate, i.e. the fractures which can generate the methane release.

Presented model of the suppression of hydrate decomposition by means of local ice jams compares well with qualitative observations of granular hydrate samples delivered to the ship board from the depth of 1400 m (Fig. 6 of Ref. (Egorov et al., 2016)). Hydrate samples on the air generate the specific click-like noise which is accompanied by periodic bounces of hydrate pieces. We suppose that such hydrate matter instability is caused by explosive exits of compressed methane from blocked fractures as the melting of the ice jams in the fractures.

# 5. Conclusions

Deep-water experiments with methane hydrates demonstrated a number of bright heat exchange phenomena, which could be predicted before theoretically but did not observed in real nature. First of all we recorded heat dissipation during transformation of deep-water methane bubbles into hydrate foam. When methane bubbles fill the trap, the maximal heat dissipation occurs on the boundary of the forming foam and the water, but heat dissipation continues in the foam due to residual water in the foam.

Alternatively, heat consumption was observed on the stage of the foam ascent with constant rate within GHSZ. The temperature in the foam was decreased by 3 $^oC$ with the decrease of the depth, i.e. the decrease of hydrostatic pressure. Video-records demonstrated that methane released from the foam sample in accordance with Boyle-Mariotte law and displaced the water from the trap. It is remarkable that heat consumption was caused exclusively by performance of the work by gas against hydrostatic pressure without any changes in foam sample.

At last, as soon as the trap crossed the top boundary of GHSZ, a temperature drop to negative value was observed. Obviously it is related with beginning of hydrate foam decomposition in the trap, which occurs with the consumption of heat. However very quickly the temperature was stabilized on the level of -0.25 $^{o}$C and did not change more up to the surface. It is noticeable, that the same temperature behavior with the same temperature stabilization on the level -0.25 $^{o}$C was observed in the similar experiment with the monolithic hydrate. Matched similarity demonstrates universal feature of effect of self-conservation which appears independently on state of hydrate, foam or monolithic sample.

Usually the effect of self-conservation is explained by the ice covering of the surface of hydrate. However our theoretical estimates show that it is not necessary to cover all surface of hydrate by the ice. It is enough to form the ice jams in most weak places of hydrate surface, namely around the fractures. Competition of formation and decomposition of the ice jams controls the maintenance of the temperature in the trap on the constant level.

## Acknowledgements

This work was supported by the Program of the Presidium of the Russian Academy of Sciences ПІ, 3П, the Fund for the Protection of Lake Baikal, and the RFBR grants 15-05-04229, 15-08-01365. The authors thank the pilot of MS 'Mir' E.S. Chernyaev for useful participation in the deep-water measurements.

## Appendix

In order to establish the possible reasons of the apperence of the minimum in dependence of the temperature in the container versus the depth, consider at a qualitative level the thermodynamics of gas expansion in the container. According to the 1st law of thermodynamics, the change in internal energy of a system d$U$, which depends only on the absolute temperature $T$, is determined by heat supply/removal $\delta Q$ and the performing of the mechanical work $p$d$v$:

$$\mathrm{d}U=\delta Q-p\mathrm{d}v$$

where $p$ is pressure, $v$ is volume of gas in the container.

Assuming that 1) internal energy of gas in the container is proportional to the absolute temperature of gas $U \propto T$, 2) the rate of supply/removal of heat is proportional to the surface area of gas bubble in the container $s$ and also to the difference of temperatures of gas and environment, i.e. $\mathrm{d}Q/\mathrm{d}t \propto -(T-T_0)s$, 3) supposing additionally that the temperature difference $\Delta T \equiv T-T_0$ is small $|\Delta T| << T_0$, we obtain finally an approximate equation of temperature variation in the container

$$\mathrm{d}\Delta T/\mathrm{d}t=-\alpha_1 s\Delta T+(\alpha_2/p)(\mathrm{d}p/\mathrm{d}t) \quad \text{(A.1)}$$

where $\alpha_1$, $\alpha_2$ are the parameters (positive), depending on the initial thermodynamic state of gas in the container, and moreover the parameter $\alpha_1$ is proportional to the thermal conductivity of the container walls, d$p$/d$t$ is rate of hydrostatic pressure drop during MS ascent.

Assuming further that MS ascent occurs at a constant speed $u_0$=-d$z$/d$t$, and at large depths the pressure varies as $p=\rho gz$, we obtain from equation (A.1) the equation of change of temperature with depth

$$d\Delta T/dz=(\alpha_1/u_0)s\Delta T+\alpha_2/z \qquad \text{(A.2)}$$

The first term on the right side of the equation (A.2) describes the temperature change due to heat exchange with the environment, the second - due to performing of the work by gas. By increasing the speed of ascent $u_0$ the influence of heat exchange is reduced, and gas is cooled more. Boundary conditions, obviously, can be presented in the form: $z=z_0$, $\Delta T=0$, $s=s_0$, $v=v_0$,, where $z_0$ is water depth of the start of the container ascent, $s_0$ and $v_0$ are the surface and volume of gas bubble in the container at this moment of time.

For qualitative analysis of changes in temperature of gas in the container, we introduce dimensionless variables: $Z=z/z_0$, $S=s/s_0$, $V=v/v_0$, $\Phi=\Delta T/\alpha_2$, which transform the equation (A.2) to the form:

$$d\Phi/dZ=BS\Phi+1/Z \qquad \text{(A.3)}$$

Accordingly boundary conditions become $\Phi=0$, $Z=1$, $S=1$, $V=1$. The parameter $B=z_0s_0\alpha_1/u_0$ in the equation (A.3) is proportional to the thermal conductivity at the boundary of gas volume and inversely proportional to the rate of ascent.

The equation (A.3) can be solved if there is a connection of gas surface area $S$ with the coordinate $Z$, $S(Z)$. It was found that the dependence of volume $V$ versus the coordinate $Z$, $V(Z)$, is described by Boyle–Mariotte law (Egorov et al., 2014). In its turn the variation of the surface area $S$ with gas volume $V$, $S(V)$, depends on the shape of container. For example, if a thin layer of gas fills a flat container, such as container 'TV-box', then variation of the volume $V$ does not change area $S\approx1$. In the case of expansion of spherical volume the surface varies with volume as $S=V^{2/3}$, whereas gas expansion in the thin vertical tube, such as 'Thermo' container, obeys to the linear relation $S=V$. According to Boyle–Mariotte law the volume of gas is inversely proportional to the depth, i.e. $V=Z^{-1}$. Thus for the above mentioned 3 cases, we have the following variations of the surface $S$ versus depth $Z$: $S=1$, $S=Z^{2/3}$ and $S=Z^{1}$. Figure A.1 shows the solution of the equation (A.3) for $S=1$, $S=Z^{1}$, $S=Z^{2}$ and discussed in the Section 4.2 the case of change of gas bubble in the container area 'Thermo' $S=Z^{1}+109.4(1-Z)^{9.6088}$. Graphs Φ(Z) show that while the area of gas bubble in the container does not grow too rapidly with the decreasing of a depth ($S=1$ or $S=Z^{1}$), temperature in the container falls monotonically. The transition to the variants of more rapid growth of the bubble surface $S$ with the decreasing of a depth $Z$, for example $S=Z^{2}$, leads to the appearance of minimum in the temperature dependence Φ(Z). The physical explanation of the appearance of minimum - is simple: at a fast increase in the bubble surface heat exchange with the environment becomes higher, so the temperature in the container tends to equilibrium with the ambient temperature and heat removal due to the expansion of gas cannot compete with heat flux from the outside.

# FIGURES

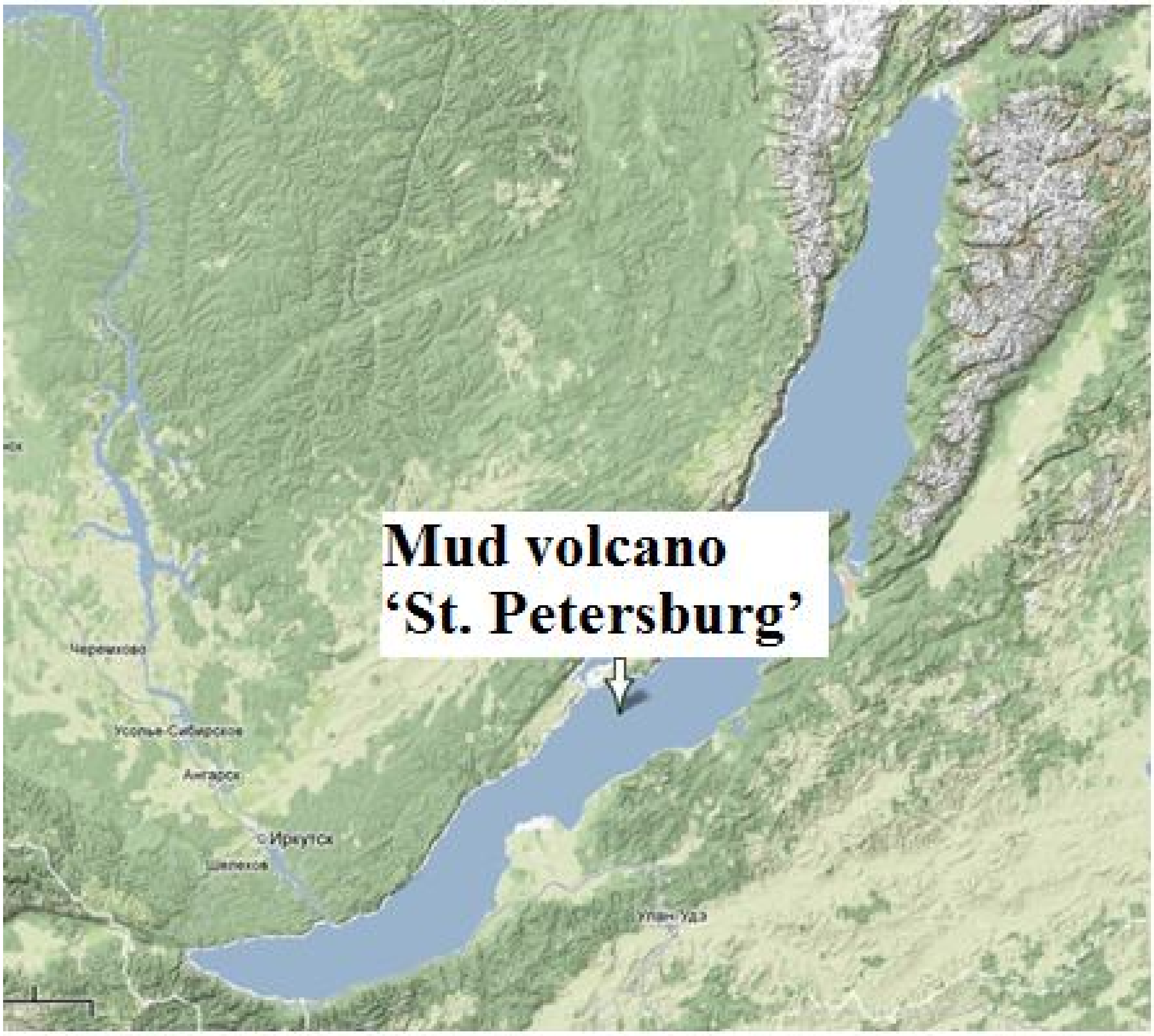

**Fig. 1.** St. Petersburg mud volcano area.

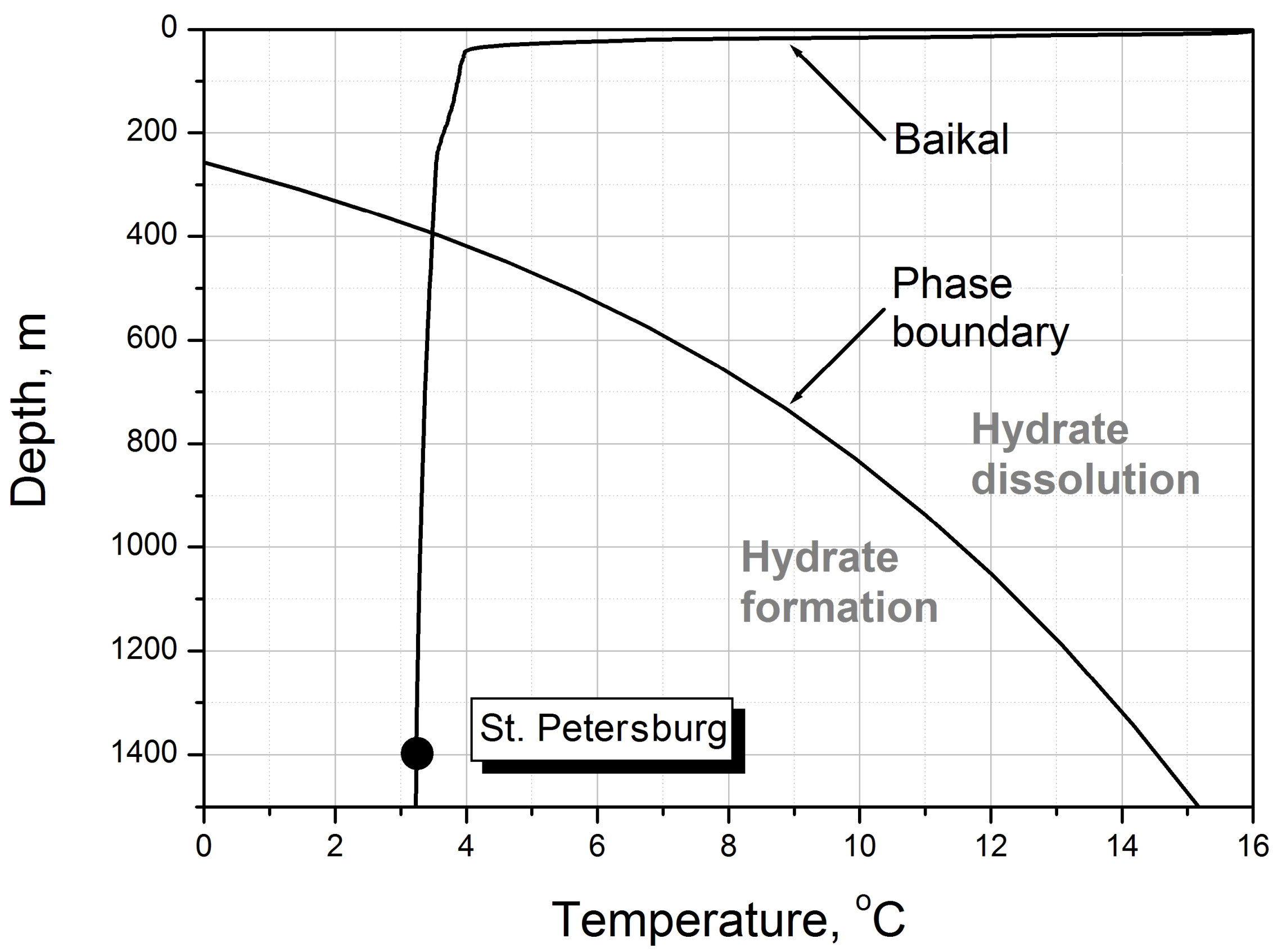

**Fig. 2.** Summer temperature profile in Lake Baikal and gas hydrate stability diagram (Katz et al. 1959). The temperature and hydrostatic pressure at the volcano depth are favorable for hydrate formation.

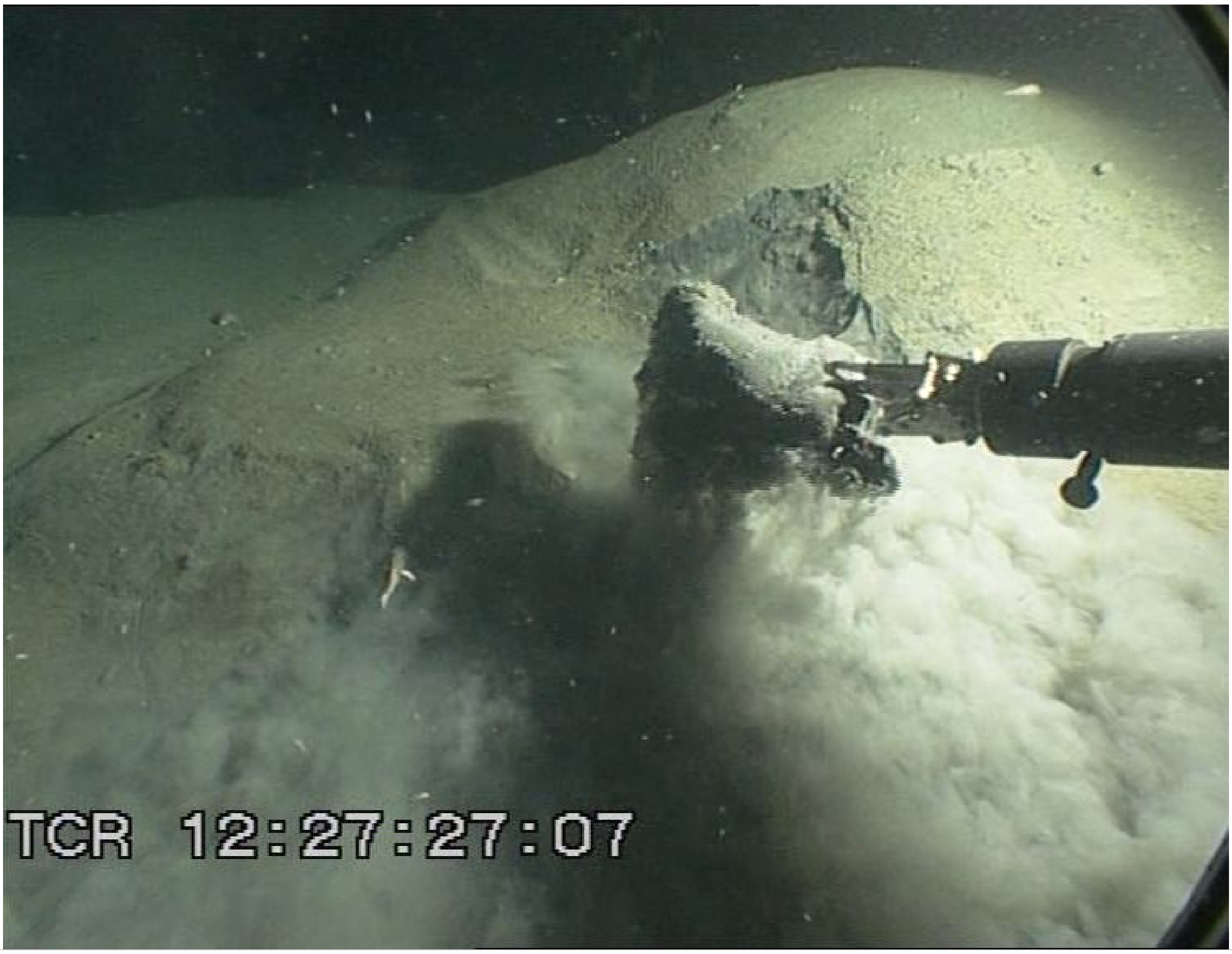

**Fig. 3.** Hydrate hills and excavation of methane hydrate sample from hydrate deposit.

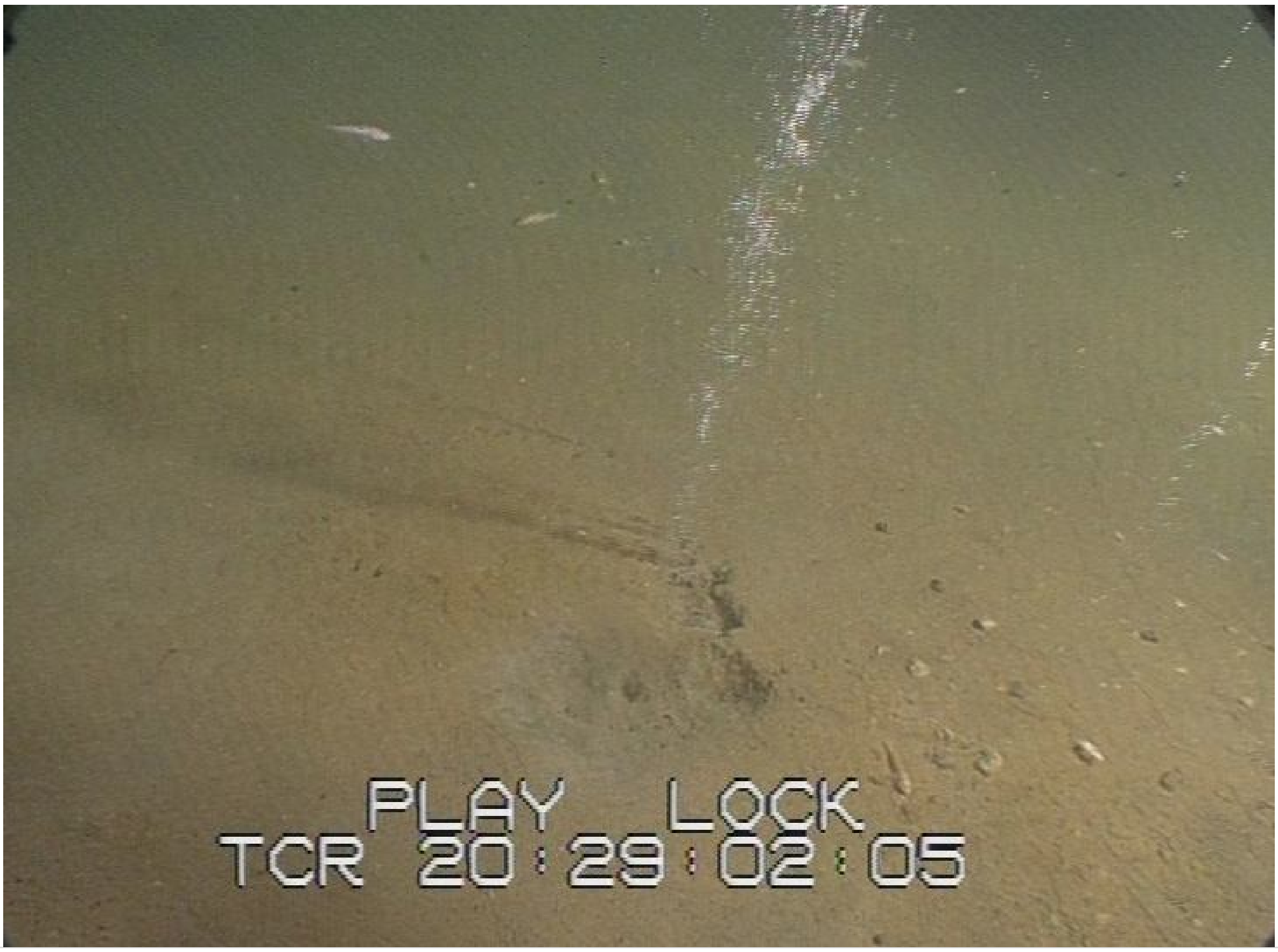

**Fig. 4.** Release of methane bubbles from deposit of methane hydrate.

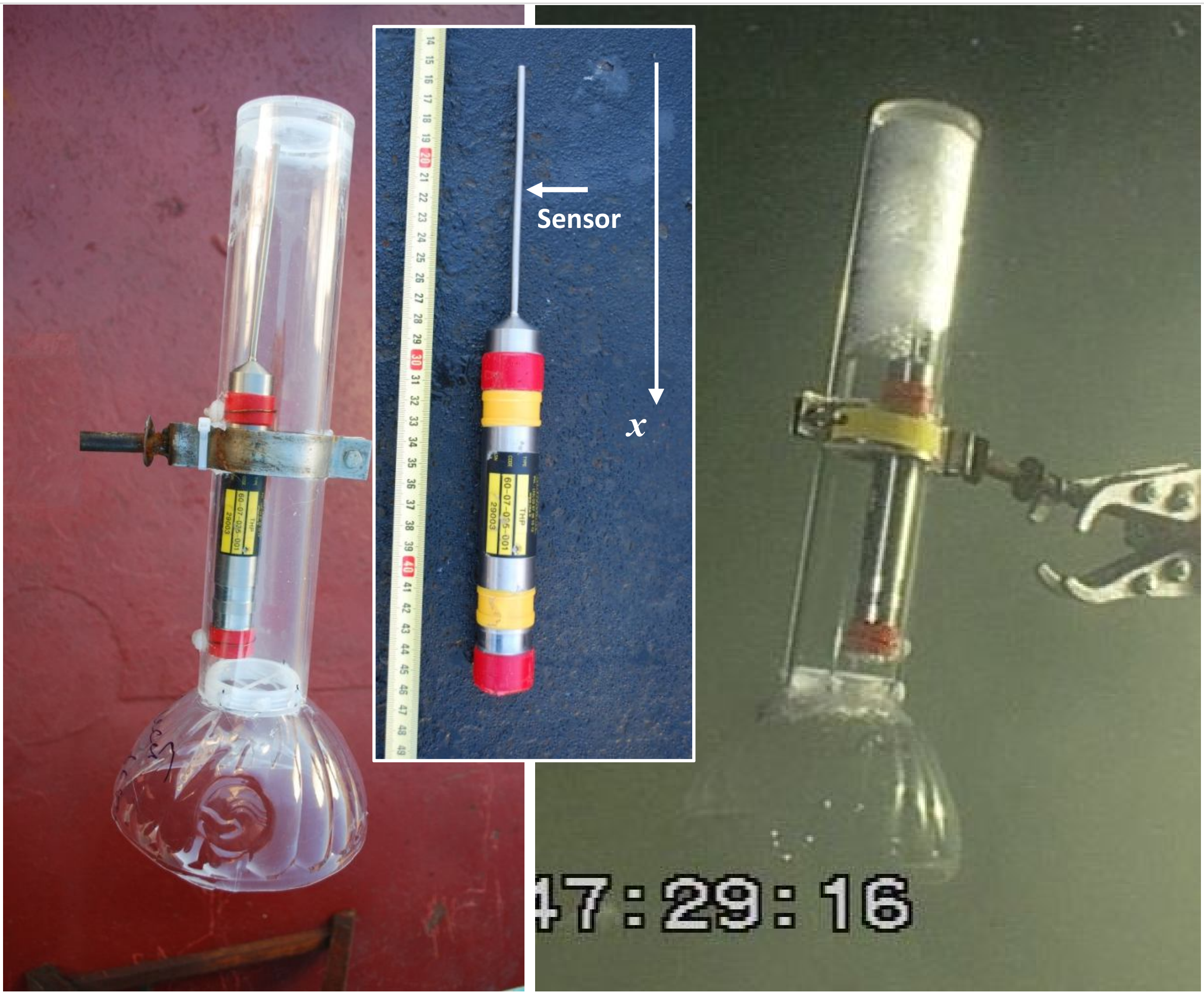

**Fig. 5.** The trap 'Thermo' before dive (left) and in the process of bubble collection near lakebed (right). Temperature sensor is located in the middle of the thermometer pin (centre).

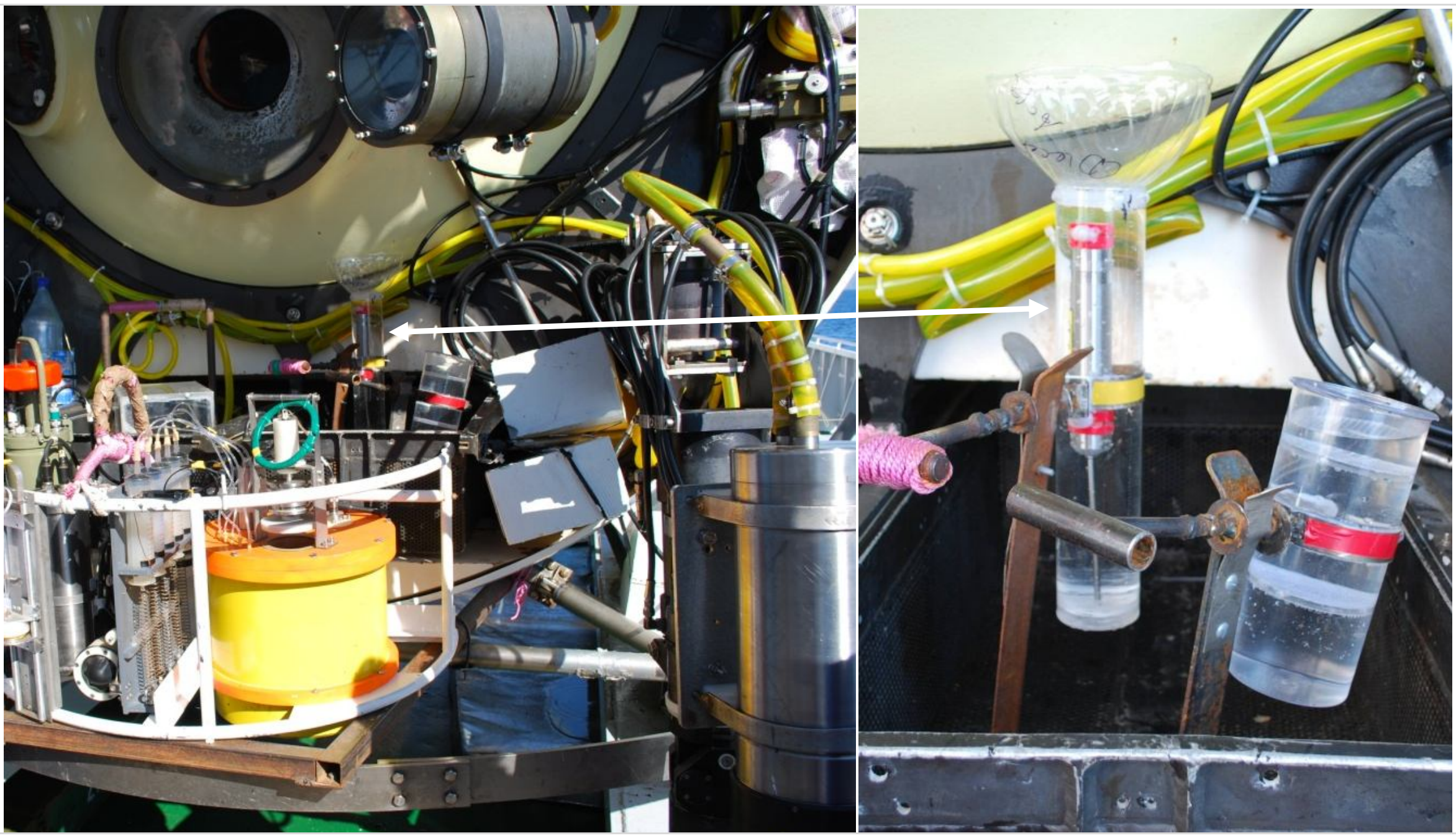

**Fig. 6.** MS 'Mir' with trap 'Thermo' just before dive. The trap is inverted and filled by water to avoid remained air bubbles inside.

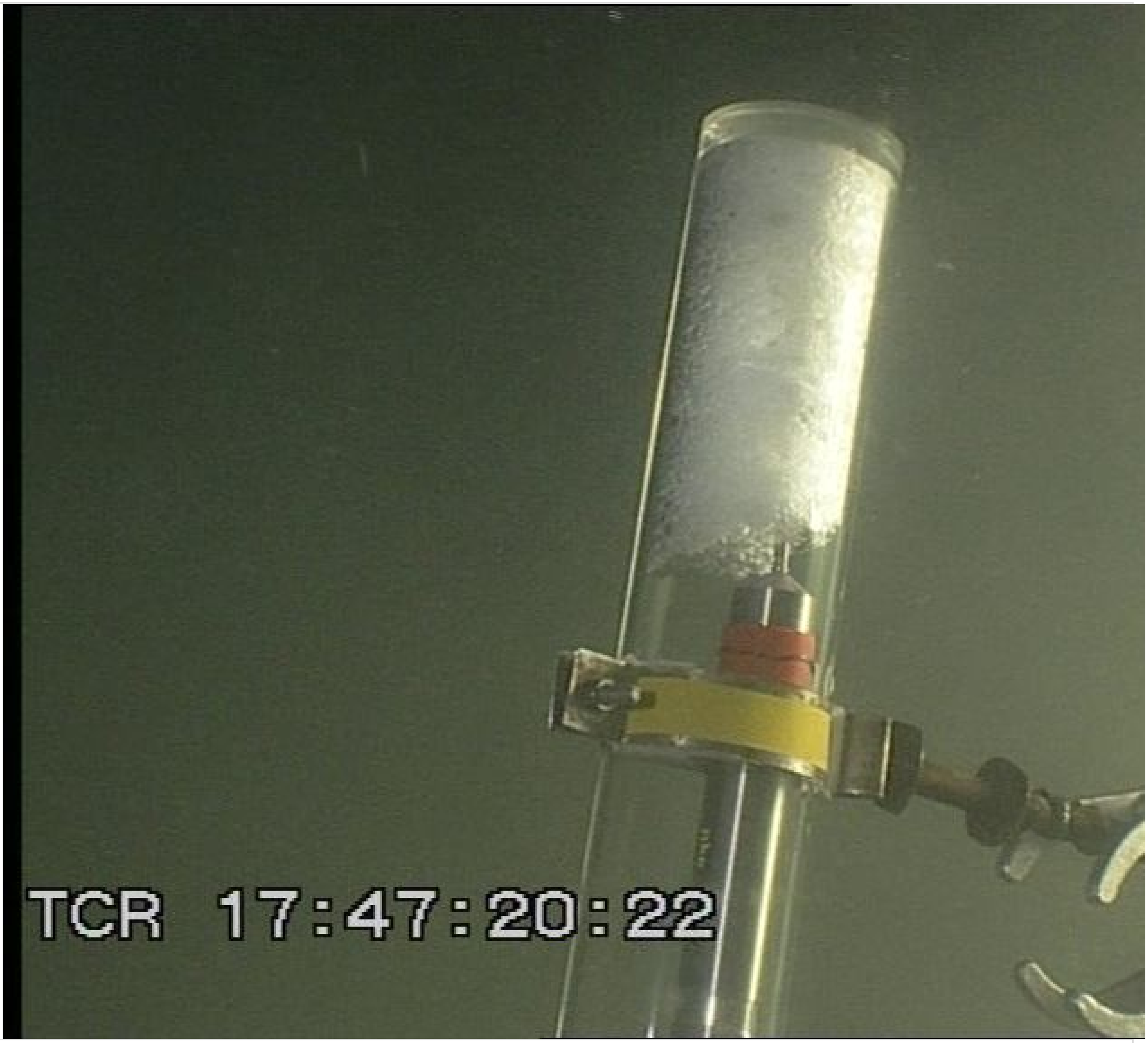

**Fig. 7.** Methane bubbles transform into column of hydrate foam.

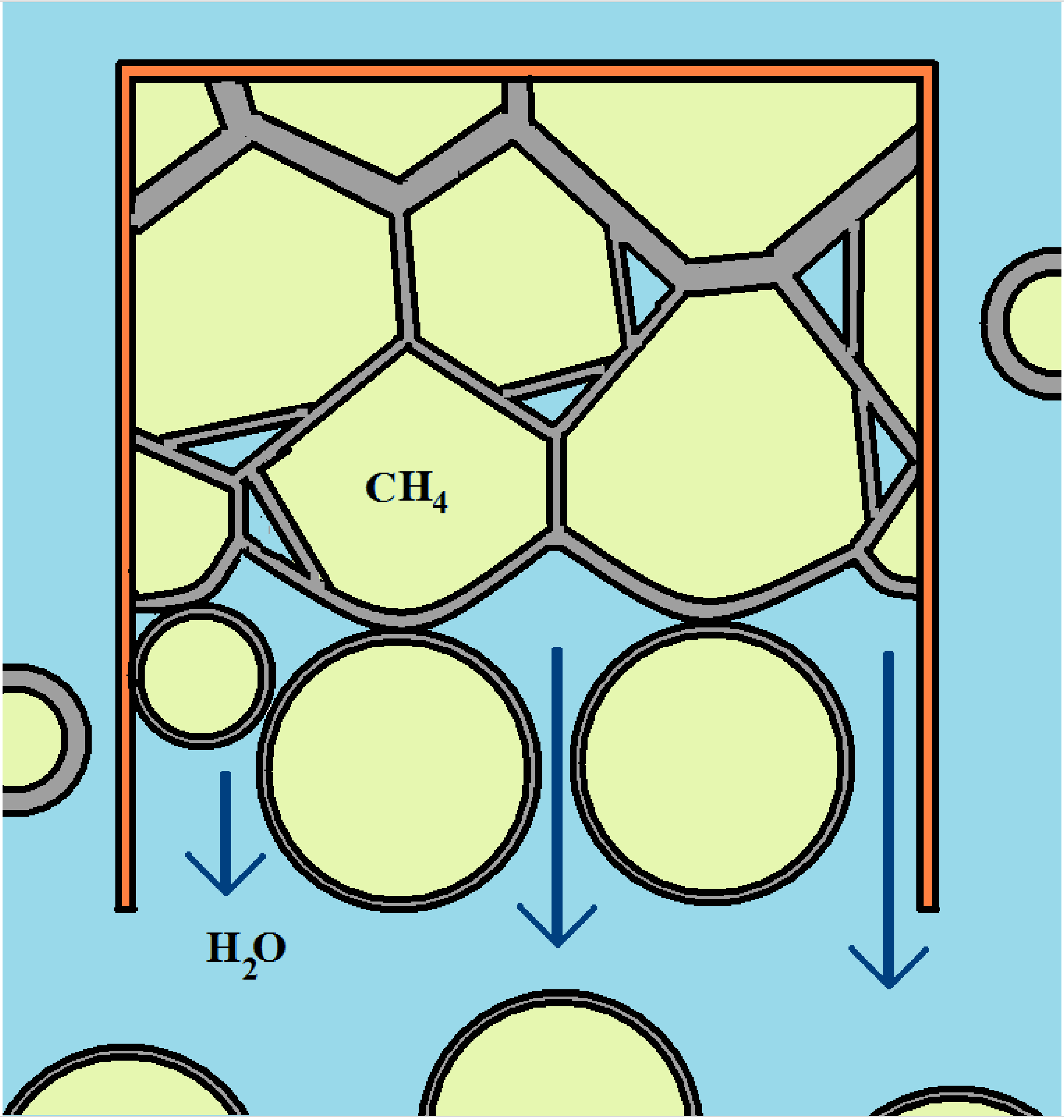

**Fig. 8.** Methane bubbles form hydrate foam in the trap. The 'islands' of water remain between methane bubbles in the foam.

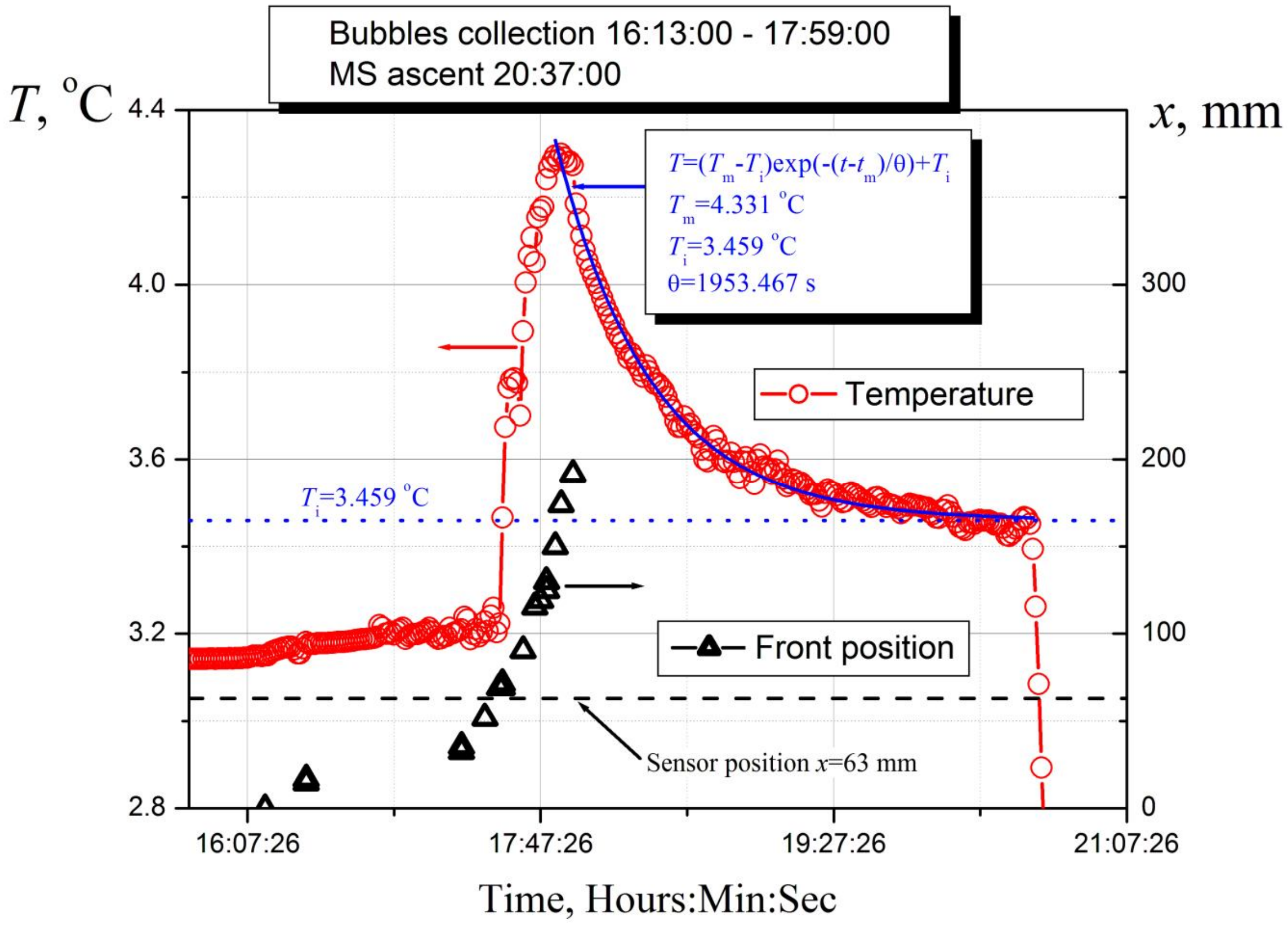

**Fig. 9.** Temperature in the trap 'Thermo' (circles, left axis) and coordinate of the front of foam formation (triangles, right axis) as functions of ship time. Dash line $x$=63 mm is sensor position. Exponential solid curve is approximation of temperature relaxation after temperature jump. Dot line $T_i$=3.459 °C shows asymptotic temperature after jump.

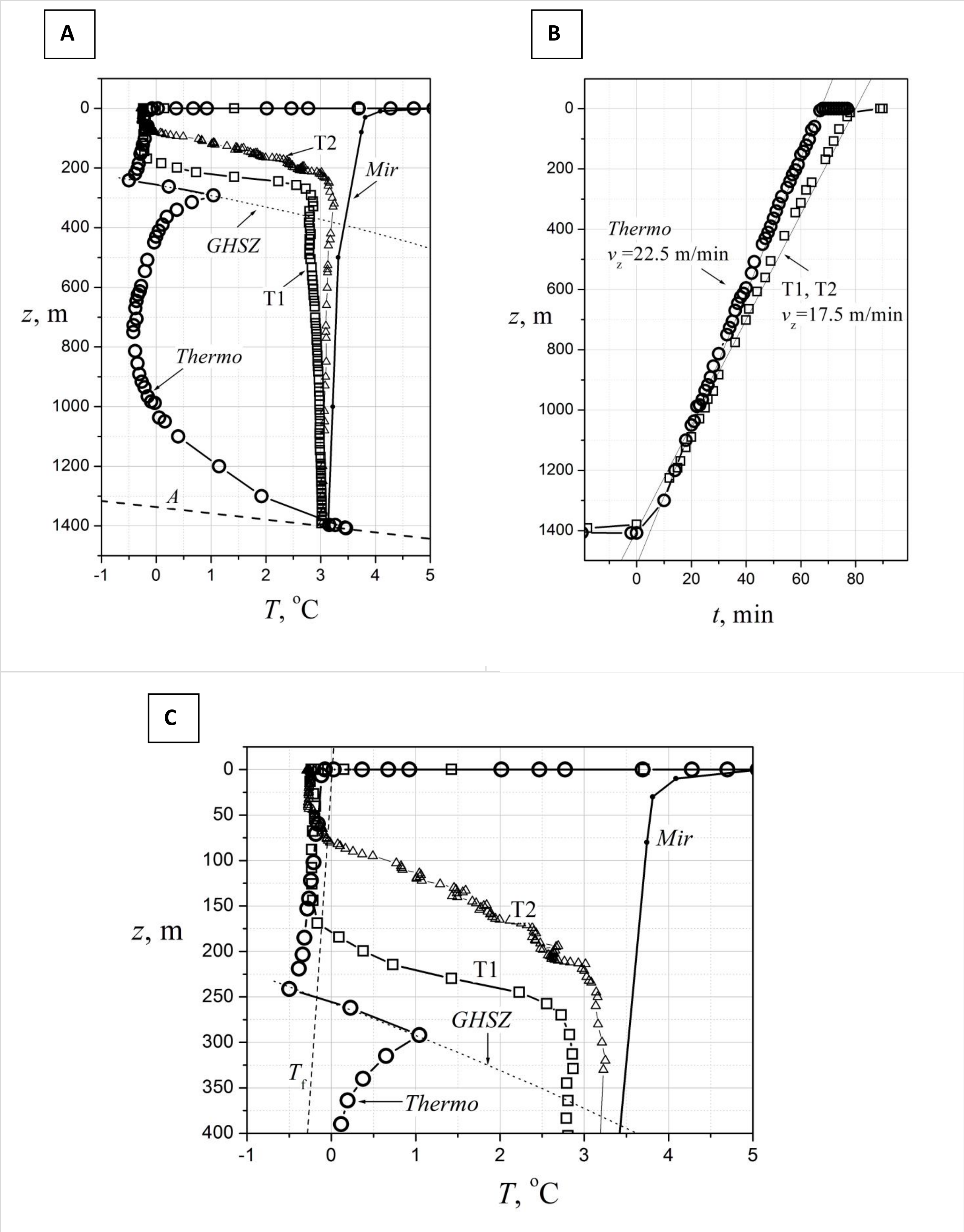

**Fig. 10. A** Temperature as function of the depth. Curve '*Thermo*' - temperature in the trap 'Thermo', 'T1', 'T2' - data of top and bottom thermometers in the container 'TV-box' (Egorov et al., 2016), curve 'Mir' - data of the thermometer of MS Mir. Curve '*A*' - the variation of the temperature with a depth in the case of adiabatic expansion of gas in the trap. Curve 'GHSZ' - top boundary of gas hydrate stability zone as function of temperature (Katz et al., 1959). **B** The trajectory of MS Mir ascent with the trap 'Thermo' and the container 'TV-box' (T1, T2). **C** Temperatures as function of a depth within GHSZ (enlarged fragment of Fig A). Curve '$T_f$' - water freezing point as function of the depth.

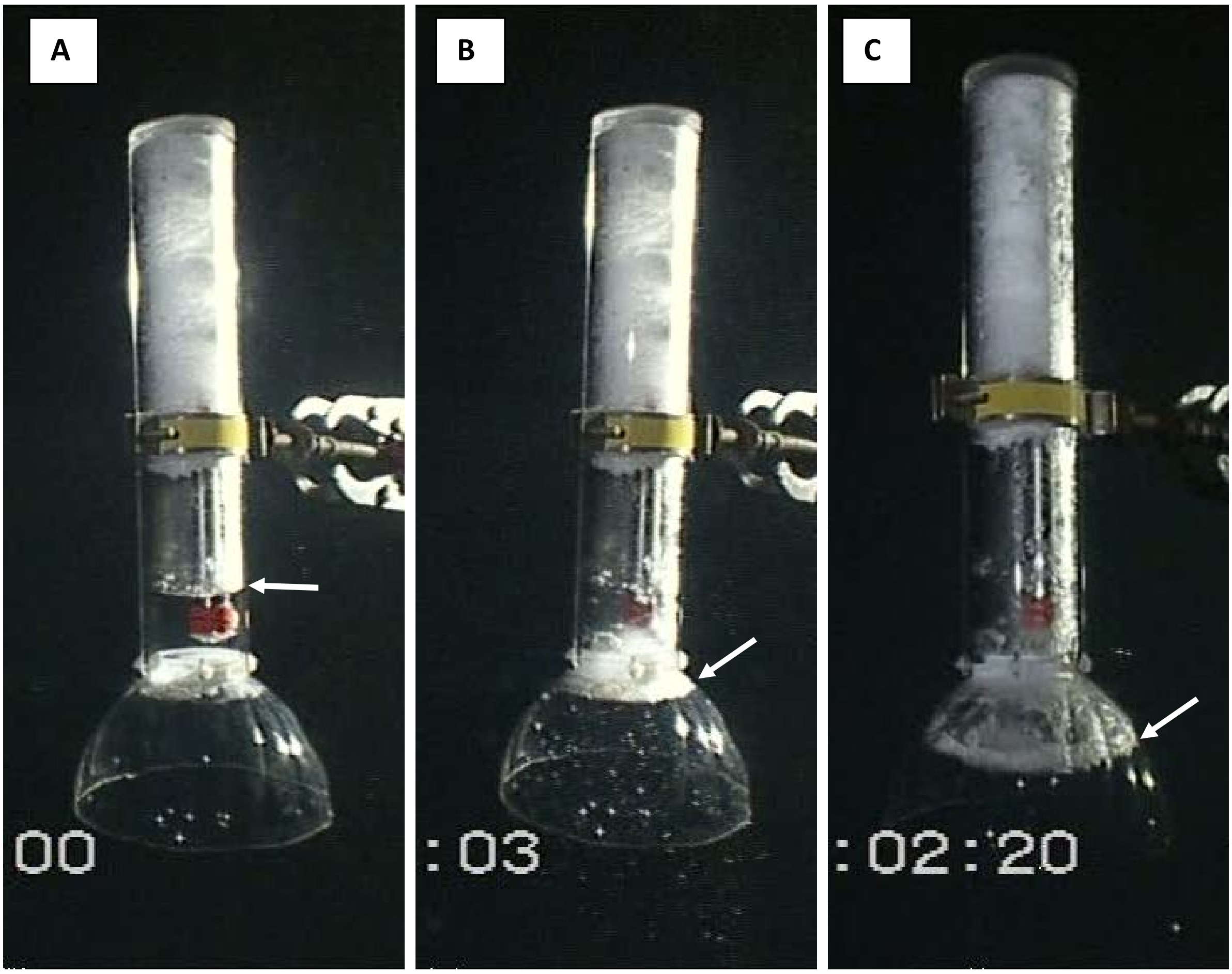

**Fig. 11.** Moving of gas-water meniscus in the trap 'Thermo' (shown by arrows) during trap ascent. Photos **A** - **B** are made at the depths of 966, 780 and 625 m. There are small hydrate particles on the surface of meniscus. By means of a slight shaking of the trap the particles can be deposited temporally into the water bulk.

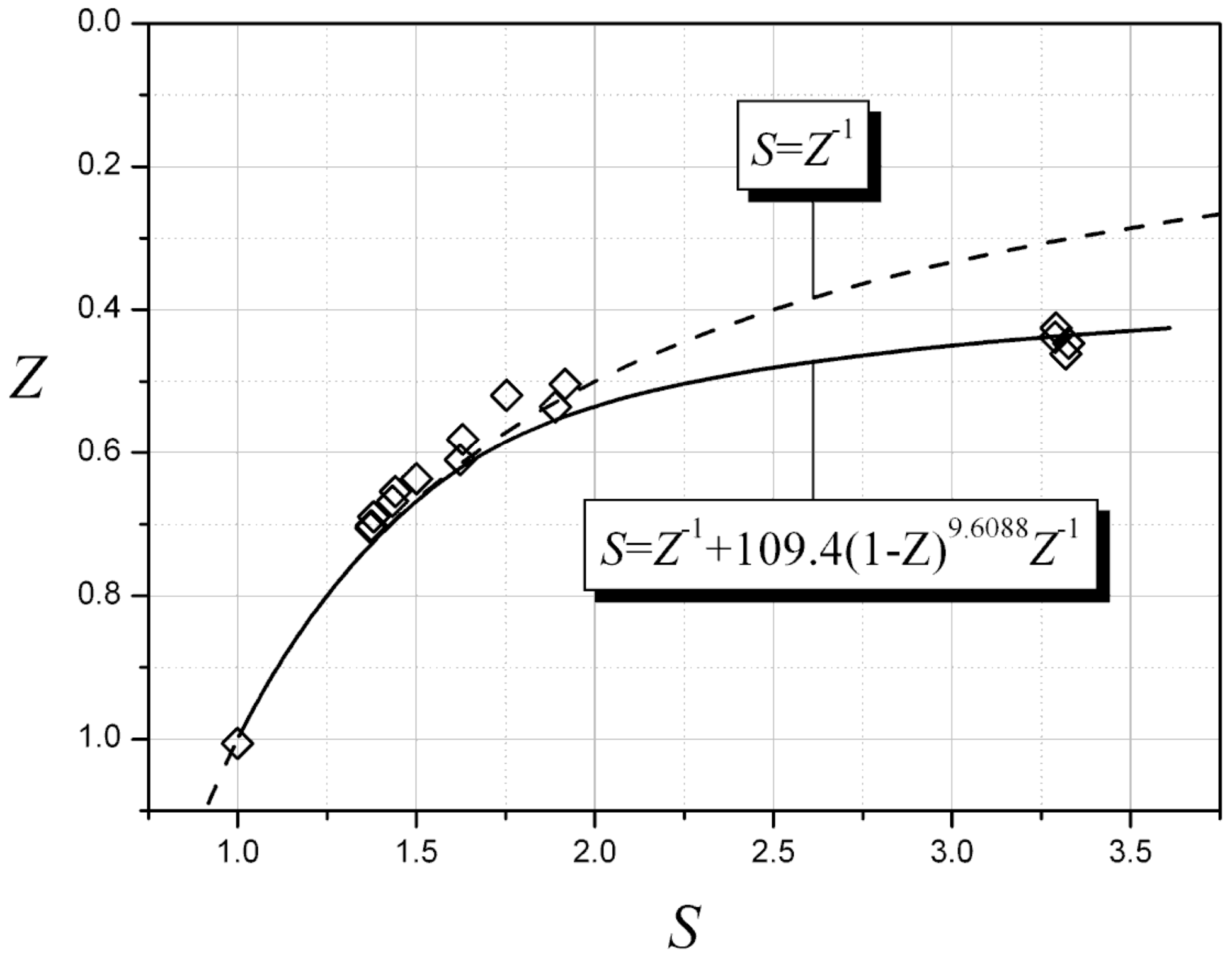

**Fig. 12.** The surface of gas bubble in the trap 'Thermo' $S$ as function of a depth $Z$. Experimental data are shown by diamonds. At a depth $Z_1$=0.5714 the meniscus moved from the cylindrical part of the trap into an expanding part (funnel).

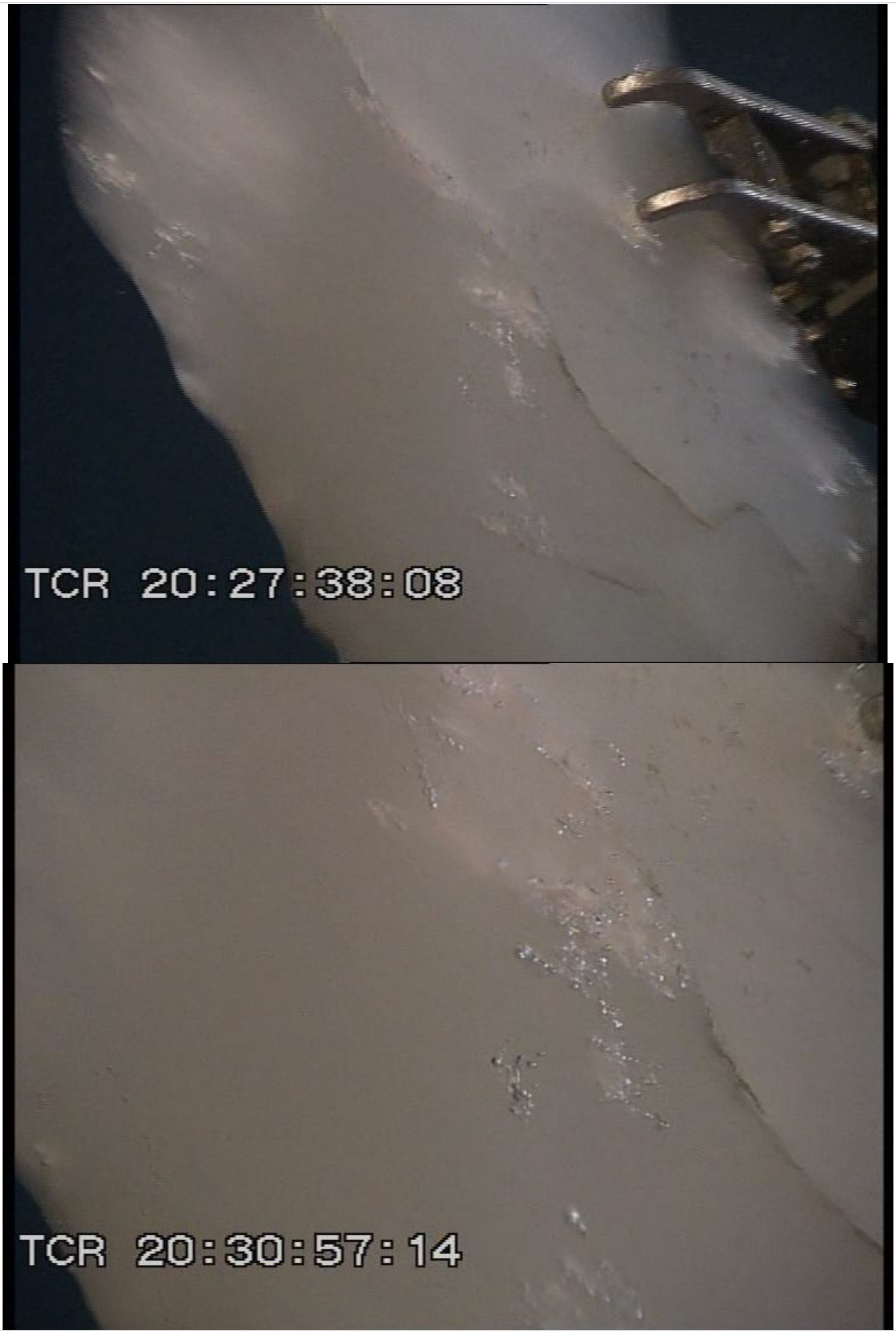

**Fig. 13.** Nucleation of micro-bubbles in the fractures in monolithic hydrate (video-records with different zoom). Hydrate is in the water slightly above GHSZ.

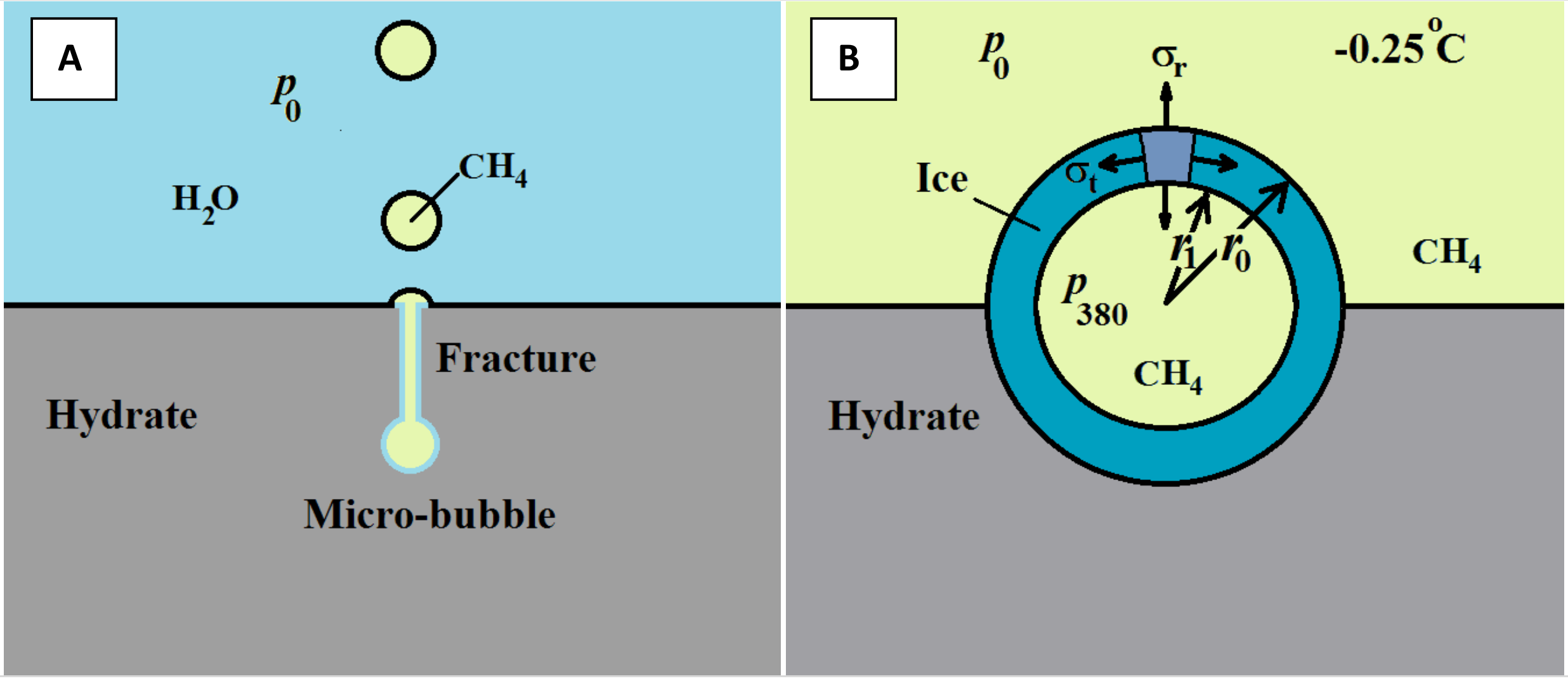

**Fig. 14.** **A** Initially hydrate decomposition occurs in the micro-bubble located in the fracture in hydrate. **B** As soon as the water is displaced by methane, the low heat exchange in methane allows the water freezing and formation of ice jam around the zone of decomposition in hydrate.

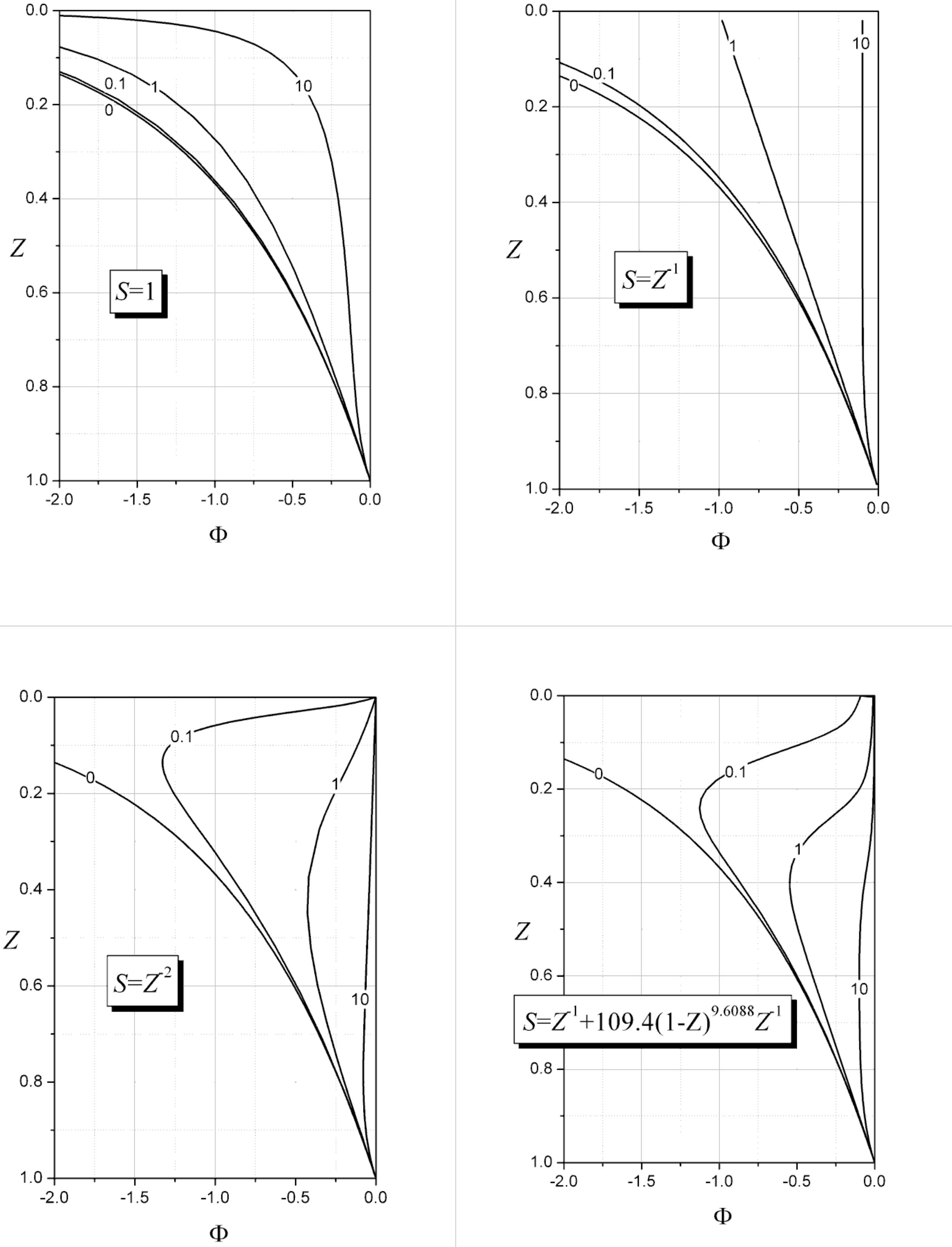

**Fig. A.1.** The solutions of Equation (A3) for different dependencies of gas bubble surface $S$ versus a depth $Z$ and different magnitudes of the parameter $B$ (shown in the graphs).